\documentclass[aps,prl,floatfix,twocolumn,superscriptaddress]{revtex4-1}
\usepackage{latexsym}
\usepackage{graphicx}
\usepackage{rotating}
\usepackage[normalem]{ulem}
\usepackage{hyperref}
\usepackage{amsmath,amssymb,amsfonts}
\usepackage{bm}
\usepackage{color}

\newcommand{\dxy}{d_{x^{2}{-}y^{2}}}

\def\onlinecite#1{\cite{#1}}

\usepackage{color}

\newcommand{\lyxmathsym}[1]{\ifmmode\begingroup\def\b@ld{bold}
  \text{\ifx\math@version\b@ld\bfseries\fi#1}\endgroup\else#1\fi}

\usepackage{amsfonts}
\usepackage{mathrsfs}

\hypersetup{colorlinks=true,urlcolor=blue,linkcolor=blue}

\begin{document}

\title{Metal-insulator transition in copper oxides induced by apex displacements}
\author{Swagata Acharya}
\affiliation{ King's College London, Theory and Simulation of Condensed Matter,
              The Strand, WC2R 2LS London, UK}
\email{swagata.acharya@kcl.ac.uk}              
\author{C\'edric Weber}
\affiliation{ King's College London, Theory and Simulation of Condensed Matter,
              The Strand, WC2R 2LS London, UK}
\email{cedric.weber@kcl.ac.uk}
\author{Evgeny Plekhanov}
\affiliation{ King's College London, Theory and Simulation of Condensed Matter,
              The Strand, WC2R 2LS London, UK}
\author{Dimitar Pashov}
\affiliation{ King's College London, Theory and Simulation of Condensed Matter,
              The Strand, WC2R 2LS London, UK}
\author{A Taraphder}
\affiliation{ Department of Physics and Centre for Theoretical Studies, Indian Institute of Technology Kharagpur, India 721302}
\author{Mark Van Schilfgaarde}
\affiliation{ King's College London, Theory and Simulation of Condensed Matter,
              The Strand, WC2R 2LS London, UK}


\maketitle

\textbf{
High temperature superconductivity has been found in many kinds of
compounds built from planes of Cu and O, separated by spacer layers.
Understanding why critical temperatures are so high has been the subject
of numerous investigations and extensive controversy.
To realize high temperature superconductivity, parent compounds 
are either hole-doped, such as {La$_{2}$CuO$_4$} (LCO) with Sr (LSCO), 
or electron doped, such as {Nd$_{2}$CuO$_4$} (NCO) with Ce (NCCO). 
In the electron doped cuprates, the antiferromagnetic phase is much more robust than the superconducting
phase. However, it was recently found that the reduction of residual
out-of-plane apical oxygens dramatically
affects the phase diagram, driving those compounds to a superconducting phase.
Here we use a recently developed first principles method to explore how
displacement of the apical oxygen (A-O) in LCO affects the 
optical gap, spin and charge susceptibilities, and superconducting order parameter.
By combining quasiparticle self-consistent GW (QS\emph{GW}) and 
dynamical mean field theory (DMFT), that LCO is a Mott insulator;
but small displacements of the apical oxygens drive the compound to a metallic state through
a localization/delocalization transition, with a concomitant maximum
$d$-wave order parameter at the transition. 
We address the question whether NCO can be seen as the
limit of LCO with large apical displacements, and elucidate the deep
physical reasons why the behaviour of NCO is so different than the hole doped
materials. We shed new light on the recent correlation observed between T$_c$ and the
charge transfer gap, while also providing a guide towards the design of
optimized high-Tc superconductors. Further our results suggest that strong correlation, enough to induce Mott gap, 
may not be a prerequisite for high-Tc superconductivity.
}


Since their discovery, the electronic structure of the high
temperature superconductors has been a subject of intensive
theoretical attention as well as controversy, a situation that
continues even today. A landmark question to understand 
these materials is how their physical properties follow from their electronic
and structural properties, and how key parameters
govern the maximum
critical temperature T$_c^\mathrm{max}$. Relating T$_c^\mathrm{max}$ to a single variable
has eluded physicists so far, because the observed correlation between T$_c^\mathrm{max}$ 
and structural properties across materials involves many kinds of them. In this work we propose
a theoretical experiment, where we study the electronic modifications induced by 
displacements of the apical oxygen (A-O), in a typical copper oxide with the T perovskite structure.
Recently, two main mechanisms have been put forward to explain the variation in T$_c^\mathrm{max}$:
i) on one hand, one body terms involving out-of-plane molecular orbitals (axial orbitals with circular symmetry in the basal plane involving
a mixture of Cu 4s, Cu 3d$_{z^2-3r^2}$, apical-oxygen 2p$_z$) \cite{pavarini_maxTc}
ii) many body effects stemming from the in-plane Mott physics \cite{cedric_tcmax,gap_trend} either in one or three band models. 
In this work, we examine this picture carefully by using a state-of-the-art approach, which treats both the many body effects and one-electron
terms on the same footing. 
We study the parent compounds La$_{2}$CuO$_{4}$
(LCO) of prototypical hole doped compounds
La$_{2-x}$Sr$_x$CuO$_{4}$. We use a realistic theoretical approach, namely
the quasi-particle self-consistent \emph{GW} theory combined with dynamical mean field theory~\cite{OLD_GABIS_REVIEW,lda_dmft_held}
(QS\emph{GW}+DMFT), as implemented in the Questaal package\cite{questaal} for QS\emph{GW}, and
Haule's implementation of DMFT \cite{Haule_long_paper_CTQMC} (see method section hereafter). 
GW and DMFT taken together
allows for a better treatment of both non-local and local fluctuations. The parent
pristine cuprate compound has for decades been
known for its complexities in both local and non-local fluctuations.
As we show, when the A-O are displaced from their equilibrium position, the role
of non-local fluctuations in spin, charge and orbital sectors is
non-trivially modified. The present scheme allows us to address many
of these entangled issues in both single and two-particle sectors.
We carefully analyze how single particle states, the bare charge transfer energy, the
correlated spectral functions, and the
optics evolve with displacement of the A-O from its equilibrium position, without distorting the otherwise octahedral geometry and
the crystal symmetry.  We represent the displacement of the
A-O along the $c$ axis by
$\delta$, such that the (Cu,A-O) bond length expands or increases from
its pristine geometry for $\delta{>}0$, and contracts for $\delta{<}0$.
We evaluate the local irreducible vertex to construct the dynamical
spin and charge susceptibilities, respectively $\chi_{s}(Q,\Omega)$ and $\chi_{c}(Q,\Omega)$ \cite{hywon_vertex}.



\begin{figure}
\begin{center}
\includegraphics[width=0.9\columnwidth]{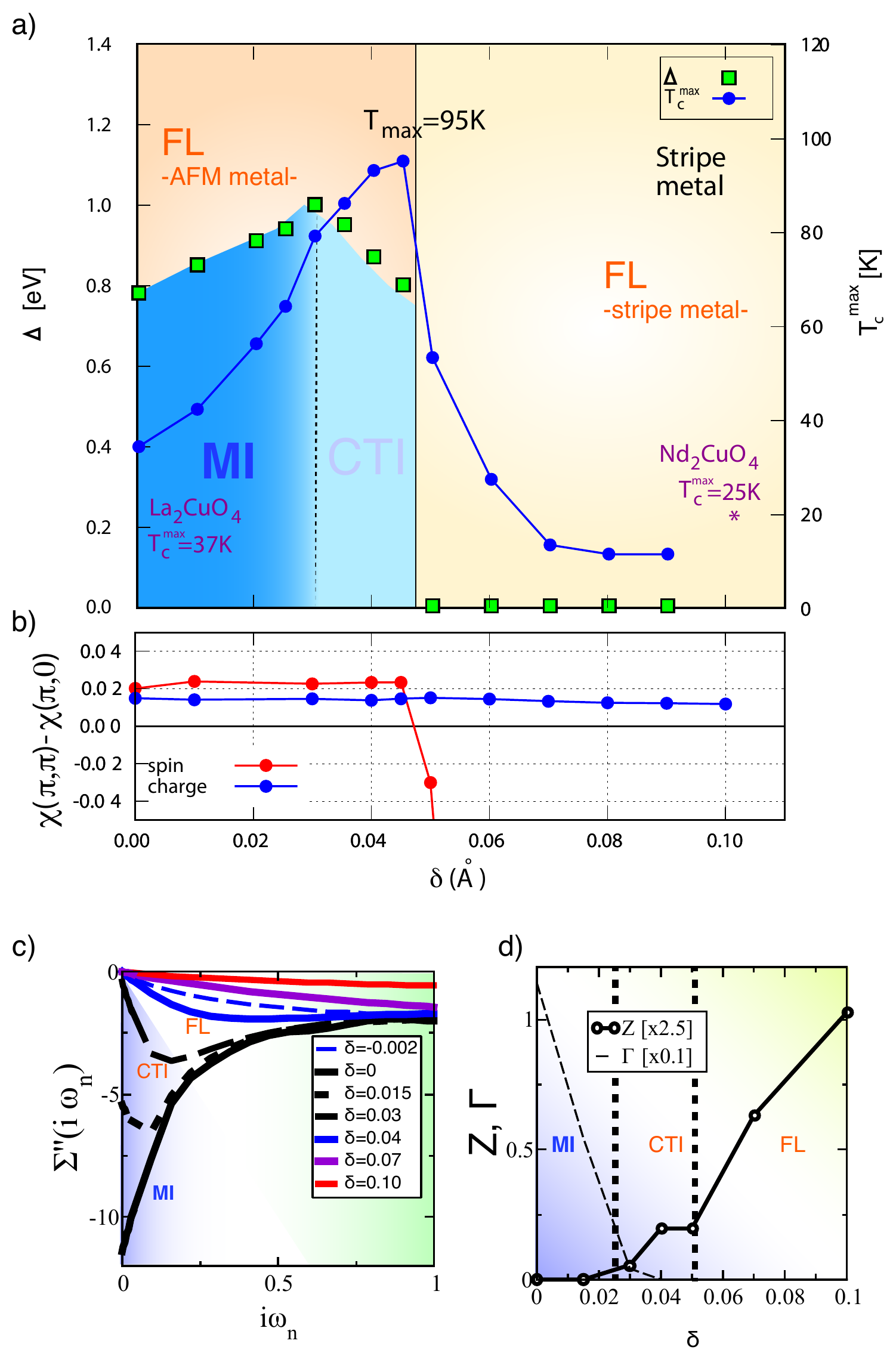}
\caption{
{\bf Phase diagram:}
(Colors online)
a) Charge gap $\Delta$ (squares, left ordinates) with respect to $\delta$, the apical oxygen displacement.
We show also the estimated critical temperature at optimal doping T$_c^\mathrm{max}$ (circles, right ordinates).
We find that optimum critical temperature is largest at the insulator to metal transition ($\delta=0.045$).
b) Relative static susceptibilities $\chi(\pi,\pi)$-$\chi(\pi,0)$ for spin (blue circles) and charge (red circles) degrees of freedom. 
c) Imaginary part of the Cu $\dxy$ self-energy $\Sigma''(i\omega)$
on the Matsubara axis for pristine LCO (bold line) and distorted LCO. For $\omega<0.03$, the self energy is extrapolated
with a polynomial fit to $\omega=0$. The large scattering rate and the pole in the self energy for $\delta<0.02$ indicate a Mott insulating phase (MI),
whereas for $\delta>0.03$ we obtain a charge transfer insulator with non-quadratic self-energies (CTI). 
d) Imaginary part of the self energy extrapolated to zero frequency $\Gamma=-\Sigma''(i\omega{\to}0)$ and quasi-particle weight Z.
}
\label{phasediag}
\end{center}
\end{figure}

Fig.~\ref{phasediag}{\bf a} shows the charge gap extracted from the spectral functions of LCO upon
displacement $\delta$.  Firstly, we find that pristine LCO is a paramagnetic Mott insulator, with a charge
gap of $\approx 0.7$\,eV. The imaginary part of the Cu ${d_{x^{2}{-}y^{2}}}$ self energy
$\Sigma_d(i\omega)$ (see Fig.~\ref{phasediag}{\bf c}) is singular
with a large degree of incoherence. 

For small, positive $\delta$ ($<0.03$\AA), the system remains a Mott insulator (MI) and we observe a 
direct correlation of the charge gap with the Cu-AO distance. This can be explained by the
reduction of the scattering
rate (see Fig.~\ref{phasediag}{\bf c})), which in turn reduces the incoherence and spread of 
the spectral weight features in the lower and upper Hubbard bands. This is also supported by the band structure
provided in the supplementary material SI (see Fig.1 of the SI). 

For $\delta>0.03$\,\AA, the self energy remains non quadratic,
but with a finite (albeit small) quasi-particle weight Z (see Fig.~\ref{phasediag}{\bf d})). The incoherence is mostly lost
and we obtain hence a crossover from a MI to a charge transfer insulator (CTI). At critical displacement $\delta_c=0.045$ the gap
collapses and we obtain a correlated metal for $\delta>0.045$\AA\ (Fermi liquid).

We now turn to the discussion of the superconducting order parameter (see Fig.~\ref{phasediag}{\bf a}). 
We estimate the d-wave superconducting pairing correlator following
the method outlined in Ref.~\cite{pairing_dwave_tremblay}. We use the magnitude of the static d-wave order parameter 
as a proxy for the maximum superconducting temperature at optimal doping (T$_c^\mathrm{max}$) \cite{cedric_tcmax}. 
Remarkably, we find that T$_c^\mathrm{max}$ \cite{cedric_tcmax} is largest at the localization/delocalization transition.
This was already obtained in model Hamiltonian studies \cite{tremblay_maxTC}. We confirm this model Hamiltonian prediction
in a realistic framework: the apex displacement provides more than a gedanken experiment, it provides 
a realization of the Mott transition in a given material with a nearly \emph{ab initio} hamiltonian, by varying a single control parameter.
Our analysis indicates that the phase transition at $\delta_c$ shapes not only the normal-state phase diagram, 
but strikingly leaves its mark on the complex structure of the superconducting condensate that is born out of this unusual normal state.
It provides hence a guide towards the design of optimized superconductors. 

Furthermore, an anti-linear correlation between T$_c^\mathrm{max}$ and the charge gap has been suggested both 
on theoretical \cite{cedric_tcmax} and experimental grounds \cite{gap_trend}. We note that 
a similar trend is obtained in our data for the CTI phase: the charge gap is reduced as the apex distance
is shifted by more than $\delta=0.03$\,\AA, but the critical temperature T$_c^\mathrm{max}$ increases up to the maximum of
${\sim}100$~K at $\delta_c$. This is understood in the picture of a loss of correlation strength, which in turn
increases T$_c^\mathrm{max}$ in the vicinity of the delocalization transition, a many body origin.

Interestingly, for small but positive $\delta$ ($<0.03$\,\AA) the increase of T$_c^\mathrm{max}$ correlates with the Cu-AO distance.
This matches with the early prediction of O. K Andersen and collaborators \cite{pavarini_maxTc}. They concluded
that T$_c^\mathrm{max}$ is controlled by the energy of the axial orbital, a hybrid between Cu 4s, A-O 2p$_z$, and Cu 3d orbitals. Our results provide a similar trend with the Cu-AO distance for $\delta{<}0.03$. Indeed, for pristine LCO and for small displacements ($\delta<0.03$\AA), the QS\emph{GW} band structure 
 indicates that the lower Hubbard d band of pristine LCO, a hybrid between 
 Cu-d$_{x^2-y^2}$ and in-plane oxygens, also has $\approx 10\%$ of axial character (see Fig. 1a and the table of the supplementary materials). The formation of the axial orbital stems from one-body part of the Hamiltonian and it is present in both DFT and GW, however, the non-local fluctuations in GW suppress the axial orbital contribution at the Fermi level significantly.

Finally, we note that the phase diagram is extremely asymmetric in $\delta$. We
obtain insulating DMFT solutions for compressions of the AO up to $\delta{=}-0.05$ and the  
charge gap increases monotonically for all such displacements. This is explained by an increased 
charge-transfer energy scale for Cu-$d_{x^2-y^2}$ and O-$p_{x,y}$.



\begin{figure}
\begin{center}
\includegraphics[width=0.9\columnwidth]{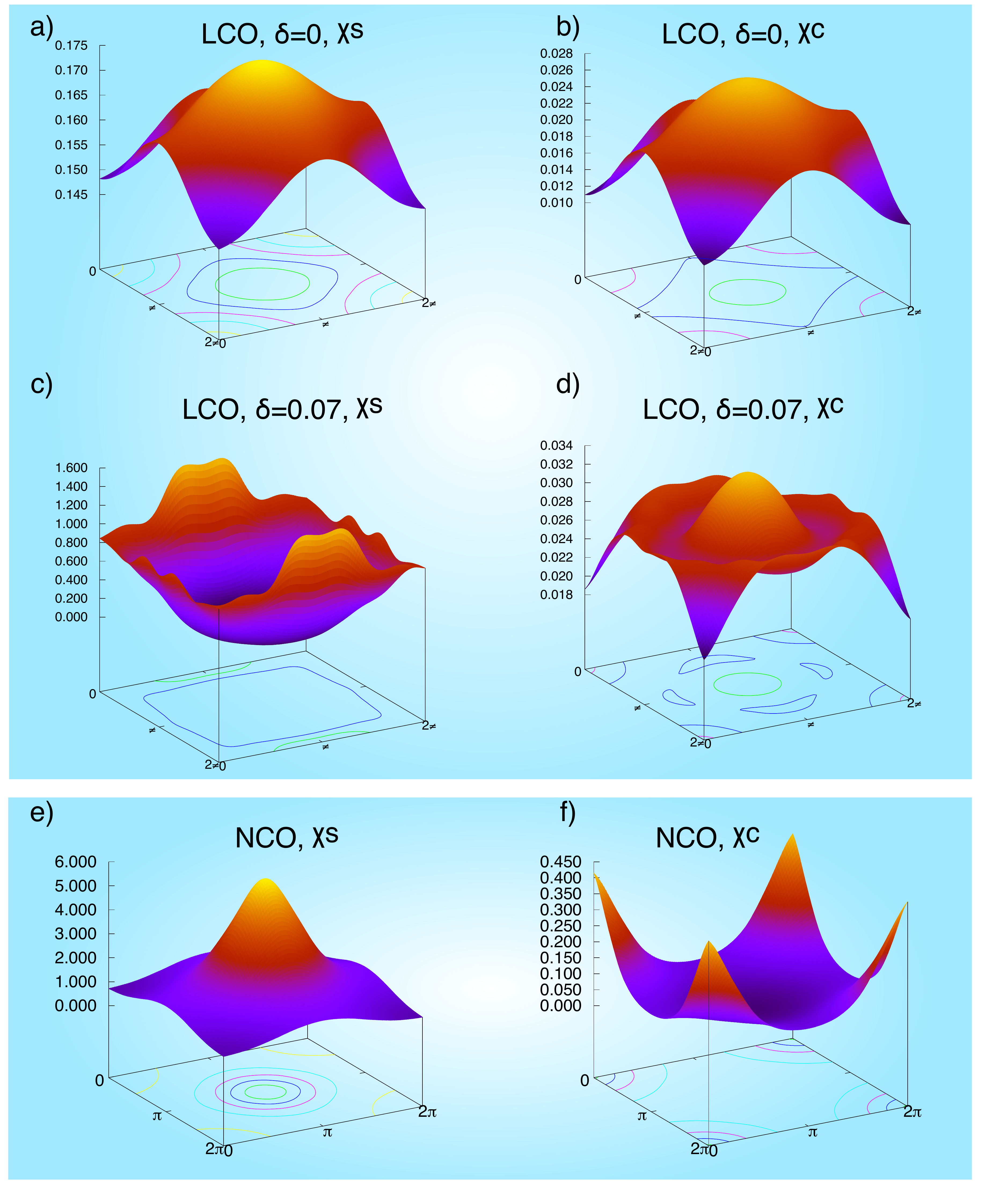} %
\caption{
{\bf Spin and charge fluctuations:} Static spin fluctuations $X^s(k,\omega=0)$ 
for a) pristine LCO, c) distorted LCO ($\delta=0.07$) and e) NCO.
Static charge fluctuations $X^c(k,\omega=0)$ 
for b) pristine LCO, d) distorted LCO ($\delta=0.07$) and f) NCO.
}
\label{fluctuations}
\end{center}
\end{figure}

Remarkably, at the critical displacement $\delta_c$, we observe a concomitant change of momentum in 
the spin fluctuations associated with the gap collapse. We show in Fig.~\ref{fluctuations}{\bf a} 
the momentum resolved static magnetic susceptibility for pristine LCO.
The susceptibility is peaked at $\bold Q=(\pi,\pi)$, but with a broad width, 
denoting short-range antiferromagnetic super-exchange $J$ in direct space. 
For displacements larger than $\delta_c$, we observe a clear shift of the peak to
$\bold Q=(\pi,0)$ (Fig.~\ref{fluctuations}{\bf c}). 

The relative static susceptibilities $\chi(\pi,\pi)-\chi(\pi,0)$ provide an order
parameter to track the change of magnetic pitch vector (see
Fig.~\ref{phasediag}{\bf b}).  While the spin fluctuations are located at $\bold
Q=(\pi,\pi)$ for pristine LCO and $\delta<\delta_c$, they abruptly evolve to
$\bold Q=(\pi,0)$ after the transition (see Fig.~\ref{phasediag}{\bf b}).

This is reminiscent of finite asymmetry present
in the orthorhombic structure of pristine LCO \cite{structureLA214}. Since the
orthorhombic phase does not respect the $x-y$ rotational invariance, the peak is
larger at ($0,\pi$) than ($\pi,0$).  As the incoherence is lost, and the
quasi-particle weight increases, the spectral features near the gap edge recover
the small structural asymmetry (for small Z the spectral feature in the lower
band are flat), and in turn provides a magnetic transition from
antiferromagnetic to a striped metallic phase.  Interestingly, the charge
fluctuations remain peaked for LCO at all time at $\bold Q=(\pi,\pi)$ (see
Fig.~\ref{fluctuations}{\bf b,d}), and they remain small in amplitude throughout
the transition.

Under compression ($\delta < 0$), the spin susceptibility
peak survives at $Q{=}(\pi,\pi)$, suggesting that there is no significant
change in the nesting and AFM fluctuations. 

Once the AO is pushed far from Cu ($\delta > 0$), however, a localization/delocalization transition with a concomitant
magnetic transition from AFM to a stripe phase, originates from the change in the shape and strength
of the spin susceptiblity.  The latter drives charge fluctuations, which reduce the local
moment and the gap.



\begin{figure}
\begin{center}
\includegraphics[angle=0,width=1\columnwidth]{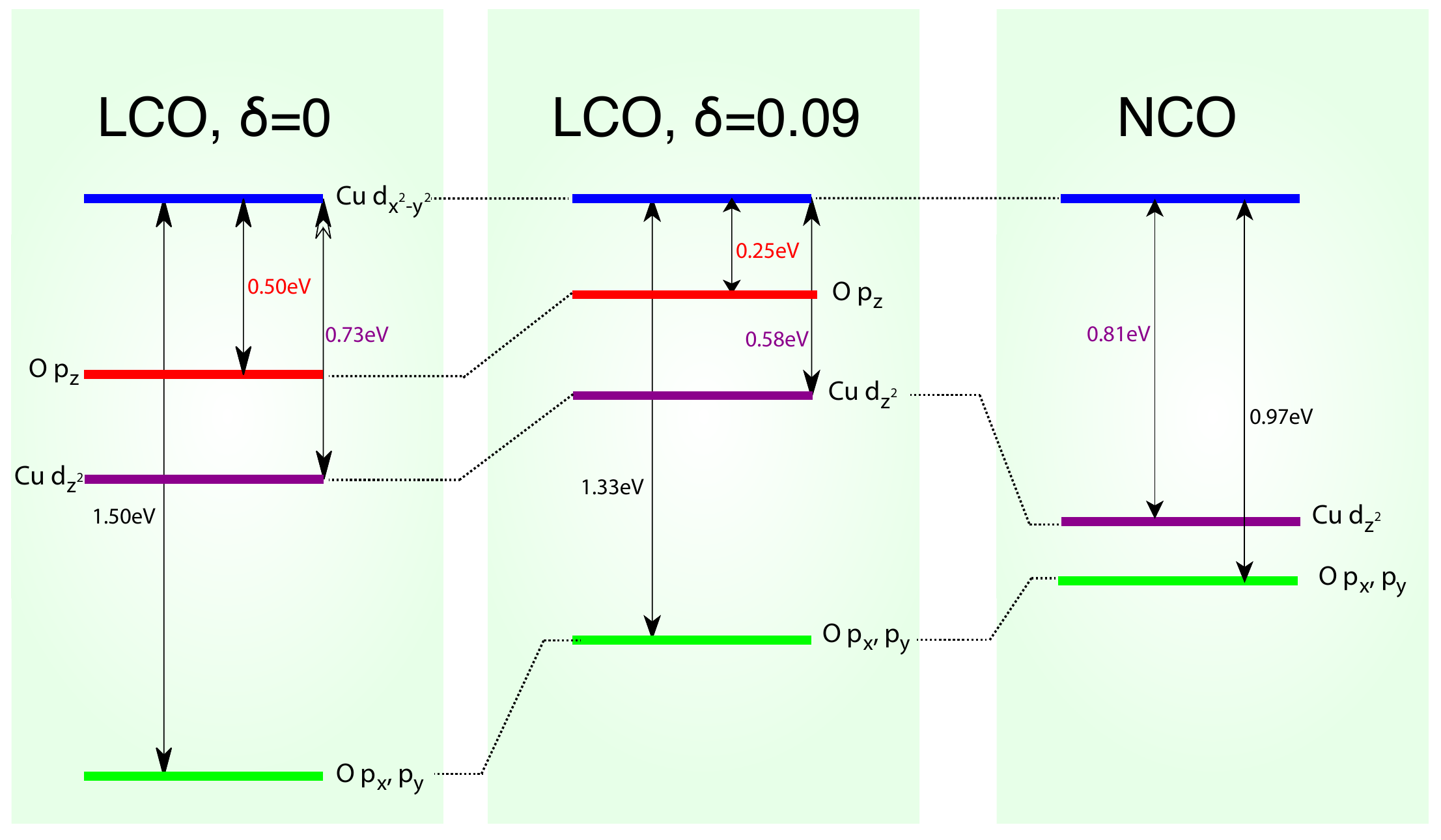}\\
\includegraphics[angle=-90,width=1\columnwidth]{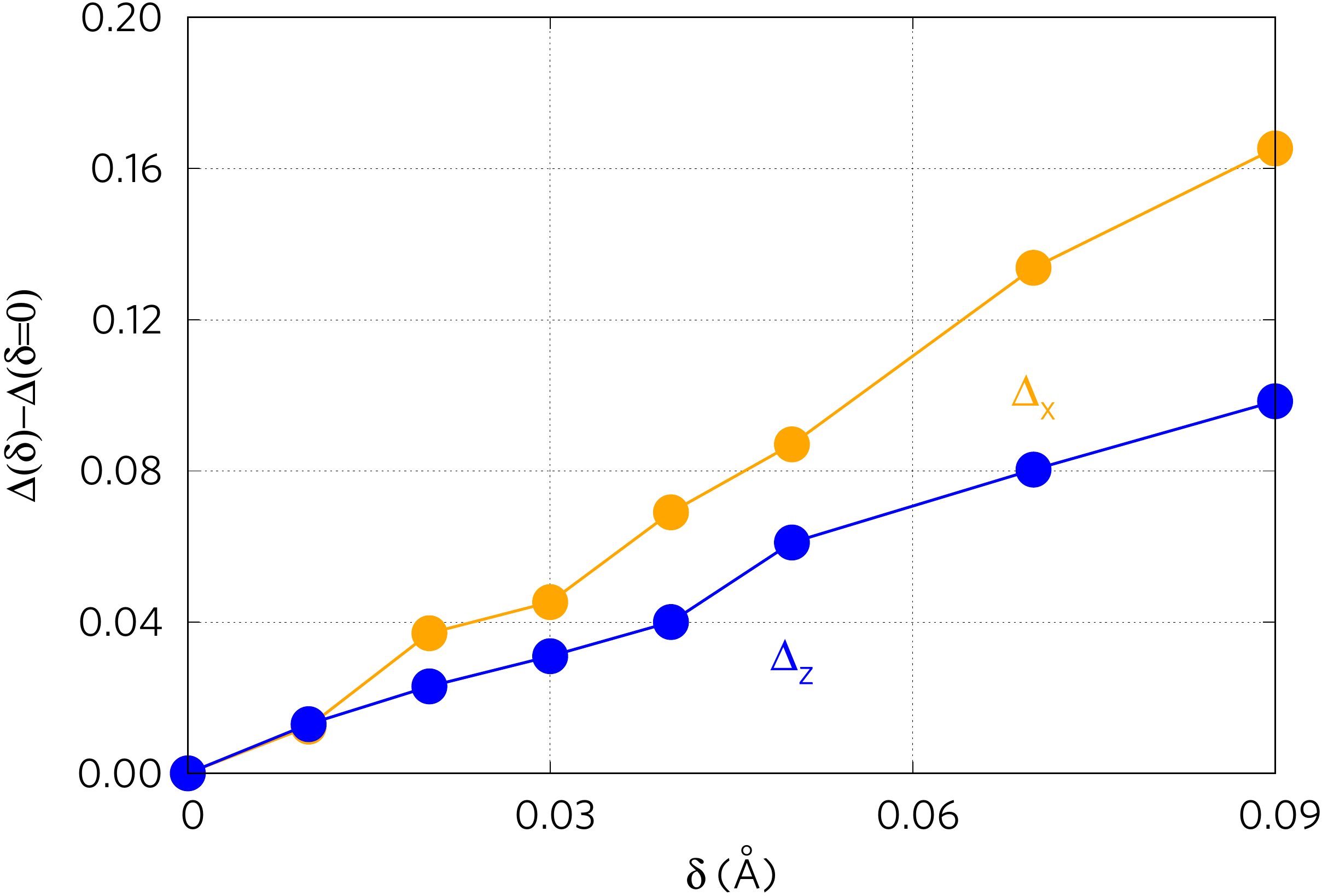}
\caption{
{\bf Energy levels and charge transfer energy:}
(Colors online)
The relative positions of the relevant QS\emph{GW} orbitals for $\delta = 0, 0.09$ and NCO (upper panel). 
Variation upon apex displacement of the in-plane charge
transfer energy $\Delta_x=\epsilon_{p_x}-\epsilon_{d_{x^2-y^2}}$ and of the out-of-plane Cu-AO charge transfer 
energy $\Delta_z=\epsilon_{p_z}-\epsilon_{d_{3z^2-r^2}}$ (lower panel). 
}
\label{chargetransfer}
\end{center}
\end{figure}

The pioneering work of Zaanen Sawatzky Allen \cite{zsa} and
their sharp (so called ZSA) boundary between charge transfer metals and
charge transfer insulators is the canonical picture to describe the copper oxides. 
The variation of the charge transfer gap (see Fig.~\ref{phasediag}{\bf a} ) suggests naively that
the charge transfer energy between the in-plane Cu 3d and the in-plane
oxygens is affected, as it is believed to determine the strength of correlations in a
charge transfer system.  

It would be expected that the Cu-AO distance would directly affect the
in-plane charge transfer energy: the
negatively charged apical oxygen increases the electrostatic
potential at the copper site, and as it is removed we expect a decrease
of the charge transfer energy $\Delta_x=\epsilon_{p_x}-\epsilon_{d_{x^2-y^2}}$.

However, our calculations show that the electostatic charge transfer energy
changes only by 0.2\,eV across the transition (see Fig.~\ref{chargetransfer}).
This effect is, however, sufficient to drive the compound to a metal, and hence
confirms that LCO is sitting at the boundary between a metal and insulator.

We also find that the out-of-plane Cu-AO charge transfer is changing on
an energy scale of $\approx 0.1$eV (see Fig.~\ref{chargetransfer}).  The latter stems from 
the re-hybridization between the Cu 3d
and out of plane orbitals, as surmised by O. Andersen and coworkers \cite{pavarini_maxTc}, via non-local 
correlations effects captured by the GW. Indeed, we observe that the band
at the Fermi level has $\approx 20\%$ character from the axial orbitals ($d_{z^2}$, O-$p_z$, Cu-4s) within the LDA calculations. 
The GW reduces this weight down to   $\approx 10\%$, but we observe that this remains present for the largest AO displacements. 
Interestingly, the axial character is nearly suppressed in NCO due to the absence of AO, with $d_{z^2}$, O-$p_z$ not contributing 
at all and Cu-4s contributing $\approx 3\%$. This suggests that the physics in NCO is well described 
by a three band model. We also find that NCO has a significantly reduced $\Delta_x = 0.97$~eV (see Fig.~\ref{chargetransfer}) than LCO ($\Delta_x = 1.50$~eV).



\begin{figure}
\includegraphics[width=1\columnwidth]{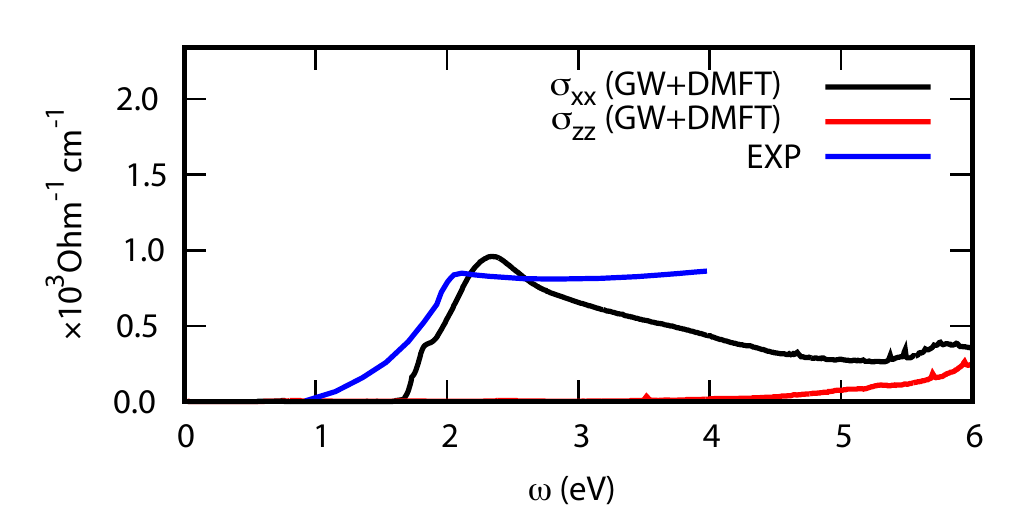}
\caption{
{\bf Optics:}
Theoretical optical conductivity $\sigma(\omega)$ obtained with the QS\emph{GW}+DMFT approach
for the in-plane ($\sigma_{xx}$) and out-of-plane components ($\sigma_{zz}$).
We emphasize that these results are a prediction from an essentially \emph{ab initio} hamiltonian.
We obtain remarkable agreement with experimental data, reproduced from Ref.~\onlinecite{optic_lco}, which
validates the theoretical approach.
}
\label{optics}
\end{figure}

We now turn to the optical conductivity (see Fig.~\ref{optics}). We find an optical gap of $\approx 1.8$\,eV with a sharp onset at around $2.0$\,eV, with $\sigma{=}920$ (ohm-cm)$^{-1}$,
followed by a plateau extending to nearly $3.5$\,eV, in remarkable agreement with 
the data extracted from Ref.~\onlinecite{optic_lco}. This provides a strong validation of the theory for pristine 
LCO, which is very important:
previous theoretical attempts had difficulty in obtaining a good description of the onset and the
plateau, and further corrections were invoked. 
In particular, DMFT calculations of the 
6-band model, including the apical oxygen and d$_{z^2-3r^2}$ orbitals, are not able
to resolve the discrepancy between theoretical optics and the experiments \cite{apical_model}.

This emerges naturally from the QS\emph{GW}+DMFT.
Indeed, the QS\emph{GW} non-local self energy aligns the O-$p$ and the Cu-$d$ states
very differently than does DFT.
There are some discrepancies with experiment: the shoulder and height of the peak
are well described, but there are difference in plateau above the shoulder.
Also the spectral weight for $\Omega{>}6$\,eV (not shown) is
blue-shifted.  Both of these effects are to be expected because of the missing
vertex in the RPA \cite{Reining}, a common problem in \emph{GW}. DMFT corrections are negligible
for $\Omega{>}3$\,eV.




Finally, performing a similar calculation for NCO, we find it to be a
correlated metal, with both $e_{g}$ and $t_{2g}$ orbitals active at
the Fermi level.  One might expect that its metallic nature would
share similarities with the limiting case of large A-O displacements
for LCO.  We find however that, despite the apparent similarities in
the one-particle response functions, the two metallic phases are
qualitatively very different, as shown by their two-particle
responses.  for instance,
we find very different magnetic (see Fig.~\ref{fluctuations}{\bf e}) and charge (see Fig.~\ref{fluctuations}{\bf f})  susceptibilities. Firstly, the charge fluctuations are peaked at 
the $\Gamma$ point, in stark contrast with LCO, where the charge fluctuations are peaked at $(\pi,\pi)$.
Second, the background of charge fluctuations in NCO are two orders of magnitude larger than for LCO.
Finally, the magnetic susceptibility is peaked at $(\pi,\pi)$, but the peak is relatively narrow, suggesting
long-range anti-ferromagnetic super-exchange interactions in direct space. 

The spin and charge separation in NCO is strongly observed in the 
susceptibilities. This suggests that the dynamics can no longer be described in terms of
individual particles or single-particle excitations, as pointed out early by P. Anderson \cite{spincharge_anderson}.
Although it is not a definite proof, our results indicate that NCO is likely to go through a 
superconducting instability, rather than through a spin- or charge-density wave alone.
Our results agree hence with the recent observation that annealed NCO and PCO compounds 
are superconducting, and not anti-ferromagnetic \cite{suppression_gap_electron_doped}.



We thank Yayu Wang for discussions and sharing insights and experimental data.
This work was supported by the Simons Many-Electron Collaboration,
and EPSRC (grants EP/M011631/1 and EP/M011038/1).
C.W. gratefully acknowledges the support of NVIDIA Corporation with the donation of the Tesla K40 GPUs used for this research,
and support from the ARCHER UK National Supercomputing Service.


\section{Method}

We use a recently developed quasi-particle self consistent \emph{GW} + dynamical
mean field theory (QS\emph{GW}+DMFT) approach to address this problem, as
implemented in the Questaal package \cite{questaal}. Paramagnetic DMFT is
combined with nonmagnetic QS\emph{GW} via local projectors of the Cu 3d states
on the Cu augmentation spheres to form the correlated subspace.  We carried out
the QS\emph{GW} calculations in the orthorhombic phase of LCO with space group
64/Cmca\cite{structureLA214}.  DMFT provides a non-perturbative treatment of the
local spin fluctuations. We use an exact hybridization expansion solver, namely
the continuous time Monte Carlo (CTQMC) \cite{Haule_long_paper_CTQMC}, to solve
the Anderson impurity problem.

The one-body part of QS\emph{GW} is performed on a $16 \times 16 \times 16$ k-mesh and charge has 
been converged up to $10^{-6}$ accuracy, while the (relatively smooth) static self-energy $\Sigma^0(\mathbf{k})$
is constructed on a $4 \times 4 \times 4$ k-mesh from the dynamical $GW$ $\Sigma(\mathbf{k},\omega)$.
$\Sigma^0(\mathbf{k})$ is iterated until convergence (RMS change in $\Sigma^0{<}10^{-5}$\,Ry).
The DMFT for the dynamical self energy is iterated, and converges in $\approx 10$ iterations. 
The calculations for the single particle response functions are performed with $10^8$ QMC steps per core and the statistics
is averaged over 64 cores. The two particle Green's functions are sampled over a larger number of
cores (192) to improve the statistical error bars. We sample the local two-particle Green's functions
with CTQMC for all the correlated orbitals and compute the local polarization
bubble to solve the inverse Bethe-Salpeter equation (BSE) for the local
irreducible vertex. Finally, we compute the non-local polarization
bubble $G(\mathbf{k},\omega)G(\mathbf{k}-\mathbf{Q},\omega-\Omega)$ and combined with the local irreducible vertex \cite{hywon_vertex} 
we obtain the full non-local spin and charge susceptibilities $\chi_{s,c}(\mathbf{Q},\Omega)$. 
The susceptibilities are computed on a $16 \times 16 \times 16$ $\mathbf{Q}$-mesh.


\section{Orbital character of Bands near the Fermi level, pristine case}

The QS\emph{GW} bands closest to the Fermi level ($E_{F}$=0 in
Fig.~\ref{spectra}), consist of mostly Cu
$d_{x^{2}{-}y^{2}}$ (red) with a substantial contribution from O $p_{x},p_{y}$ (blue) orbitals (Table I).  Thus near
$E_{F}$ bands are purple.  There is also 
a small but finite contribution from orbitals of axial symmetry
(mostly O-$p_{z}$ but also Cu-d$_{3z^2-1}$ and Cu-$4s$) on the
$A{-}E_{0}{-}T$ and $S{-}R{-}Z$ symmetry lines.  The Cu-$4s$ weight is most
pronounced near the Y point, in a band labelled Cu $s$.  O $p_{z}$ (orange,
below ${-}3$ eV) hybridise to some extent with Cu $d_{3z^2-1}$ (green)
around ${-}2$\,eV and the other Cu $d$ orbitals (black, labelled as Cu $d$),
and to a lesser extent, the bands at $E_{F}$.

The LDA bands at the Fermi level consist of strongly mixed Cu-$d_{x^{2} -
  y^{2}}$ (red) and O-$p_{x}, p_{y}$ (green) (Fig.~\ref{spectra}{\bf a})
orbitals.  The Cu-$d_{3z^2-1}$ orbitals also have strong mixing with primarily
apical O-$p_{z}$ orbital 1.9 eV below the Fermi level. The O-$p$ states are strongly hybridized with
most orbital characters throughout the entire window
$(E_{F}{-}8\mathrm{\,eV},E_F)$.  However, in contrast to the LDA, QS\emph{GW}
splits out the O-$p$ orbitals from the Cu-$d$ and puts them roughly 2\,eV
(Fig.~\ref{spectra}{\bf a}) below the (green) band with dominant Cu $d_{3z^2-1}$
character.  Also, within QS\emph{GW} the band character at the Fermi level
becomes dominantly Cu-$d_{x^{2}{-}y^{2}}$, with significantly lessened O-$p_{x},
p_{y}$ contribution than the LDA. 


Finally, the optical absorption predicted by QS\emph{GW} + paramagnetic DMFT are
in good agreement with experiment, as described in the main text.

All the relevant numbers for the contributions from different orbitals to the active band at the Fermi level are shown in the table.

\subsection{QS\emph{GW} and LDA compared}

When comparing bands of similar color in the first and second columns of
Fig.~\ref{spectra}, it is evident that there are qualitative similarities but
important quantitative differences.

\begin{itemize}
\item The ${x^{2}{-}y^{2}}$ band is about 50\% wider in the LDA, and
the asymmetry in the splitting around $E_{F}$ is inverted.

\item Contrary to what would be expected from the wider bandwidth,
the density-of-states at $E_{F}$ is about 50\% \textit{larger} in the LDA.
This indicates that not only level alignments, but hopping matrix elements
are substantially different in the LDA.

\item O $p$ (blue and orange) are 1.5-2 eV closer to $E_{F}$ in the LDA.  As a
consequence they hybridise much more strongly the Cu $d$ orbitals,
in particular the ${x^{2}{-}y^{2}}$ and ${3z^2-1}$ orbitals that
drive superconductivity.

\item Splitting between Cu ${x^{2}{-}y^{2}}$ and ${3z^2-1}$ orbitals
is roughly twice larger in the LDA.

\item Orbitals corresponding to those labelled Cu ${s}$ and La $d$ 
  in the QS\emph{GW} and LDA panel lie about 1\,eV closer to $E_{F}$ in the LDA.
  Moreover, Cu $s$ portion to the band labelled ``Cu $s$'' is about
  half of what it is in QS\emph{GW}.

\item the La $f$ states (dense black, around 3.5 eV in the LDA) should be at around 10 or 12 eV.

\end{itemize}

It is interesting to note that the non-magnetic bath once coupled with magnetic
DMFT, using 82.2 eV for double counting corresponding to 9 $d$ electrons, does
not produce a Mott gap at the Fermi level when LDA is used for the bath, while
QS\emph{GW}+paramagnetic DMFT does.  The spectral function computed from DMFT
shows the gap opening (Fig.~\ref{spectra}{\bf a}) at the Fermi level within this
formalism for the parent LCO,  while the DMFT spectral functions
(Fig.~\ref{spectra}{\bf b}) when $\delta{>}0.045$\,\AA\ has metallic excitations
across the Fermi level.

\section{Apical Oxygen Displacements and the Spectral Evolution}


Shifting the apical O has a delicate effect on the QS\emph{GW} hamiltonian;
nevertheless this small change is enough to give rise to dramatic changes when
many-body effects are included through DMFT, showing that pristine LCO is in the
vicinity of a quantum-critical point.

\begin{figure*}
\includegraphics[width=1\textwidth]{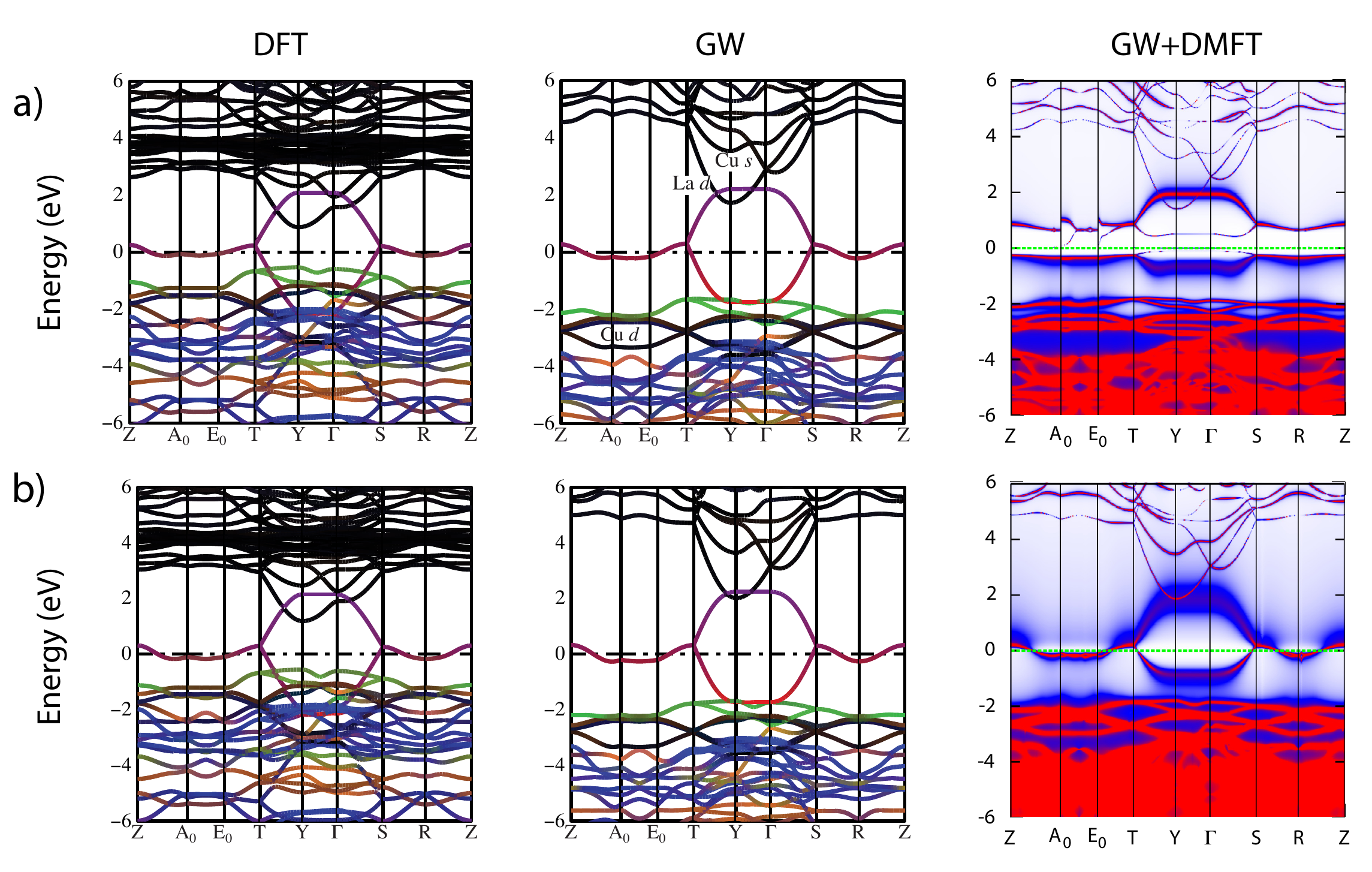}
\caption{
{\bf Band structure:}
	Comparison of the band structure as obtained from DFT, QS\emph{GW}, and
	QS\emph{GW}+DMFT. 
	a) pristine LCO ($\delta{=}0$).
        b) compound with apical oxygens displaced by $\delta{=}0.09$\,\AA.
         c) zoomed in band structure for pristine LCO and d) for distorted LCO $\delta{=}0.09$\,\AA.
	Colors reflect the orbital character of the bands as follows:
        Cu $d_{x^{2}{-}y^{2}}$ (red);
        Cu $d_{3z^2-1}$ (green), near $-$2\,eV in QS\emph{GW};
        O-$p_{xy}$ (blue), between $-$3\,eV  and $-$5\,eV in QS\emph{GW};
        O-$p_{z}$ (orange), between $-$3\,eV  and $-$5\,eV in QS\emph{GW}.
        The band near 3.3\,eV at Y has significant Cu $s$ character,
        and it also admixes  into the Cu $d_{x^{2}{-}y^{2}}$ near $E_{F}$.
}
\label{spectra}
\end{figure*}

When O is displaced by $\delta$=0.09\AA, the Cu $d_{3z^2-1}$ drops about 0.2\,eV
relative to the ${x^{2}{-}y^{2}}$ state, while the O $p_z$ rises by an
approximately similar amount.  These states come closer together, reflecting the
reduction in the hopping matrix element that hybridises them and pushes them
apart.  At the same time, the O $p_{xy}$ bands rise by $\sim$0.1-0.2\,eV.
Fig.~\ref{spectra} shows valence bands resolved on a fine energy scale for
$\delta$=0 and $\delta$=0.09\AA, where the shifts can be seen in detail.  As a
secondary effect, the Cu ${x^{2}{-}y^{2}}$ bandwidth is slightly reduced.

The lowest bands above $E_{F}$ are primarily of La character, especially La $d$.
La does not hybridize strongly with the Cu and O, but there is a significant
evolution in charge transfer to it with $\delta$.  The shift in the bands is a
consequence mainly of the difference in electrostatic potentials at the La and
Cu nuclei. In the $\delta$=0.09\AA\ structure this difference is $\sim$1.5\,eV
larger than in the pristine compound.

\newpage
\pagebreak
\begin{table*}
\begin{tabular}{ |p{4cm}|p{1cm}|p{1cm}|p{1cm}|p{1cm}|p{1cm}|p{1cm}|p{1.6cm}|p{2cm}|p{1cm}|}
	 \hline
	  \multicolumn{10}{|c|}{Table} \\
	   \hline
	     \text{Compound}  & $Z$ & $\Gamma$ & N$_{eff}$ & $\Delta_x$ in eV & T$_c^\mathrm{max}$ in K & $\Delta$ in eV & Cu $d_{x^{2}{-}y^{2}}$ &  Axial orbitals & O-$p_{xy}$\\
	     \hline
	      LCO ($\delta=0$) QS\emph{GW}  & 0.00    & 11.4 &  0.3 & -1.53  & 34 & 0.78 & 0.5437  & 0.108 & 0.2630\\
	       LCO ($\delta=0.03$) QS\emph{GW} &   0.02  & 0.41   & 3.4 & -1.49 & 79  & 1.0 & 0.5454  &  0.105  &  0.2615\\
		LCO ($\delta=0.05$) QS\emph{GW} & 0.08 & 10$^{-4}$ &  42.6 & -1.45 & 53 & 0 & 0.5471  &  0.092  &  0.2607 \\
		 LCO($\delta=0.09$) QS\emph{GW} & 0.41 & 10$^{-9}$ & 86.0 & -1.37 & 11 & 0 & 0.5507  &  0.086 &   0.2591 \\
		 LCO($\delta=0$) LDA &  & & & & & &   0.40067 &   0.22 &   0.2798 \\
		  NCO QS\emph{GW} & 0.44  & 10$^{-6}$ & 98.3 & -0.97  & 25 & 0 & 0.63007 & 0.033 & 0.2520  \\
		  \hline
		  
\end{tabular}
\caption{Quasi-particle weight Z, Imaginary part of the self energy extrapolated to zero frequency $\Gamma=-\Sigma''(i\omega{\to}0)$, optical mass $N_{eff}$ (optical conductivity integrated up to 2.3 eV), charge transfer energy $\Delta_x$, d-wave superconducting transition temperature T$_c^\mathrm{max}$, single particle charge gap $\Delta$, relative weights of the Cu $d_{x^{2}{-}y^{2}}$, axial orbitals (Cu $d_{z^{2}}$, O-$p_{z}$, Cu ${4s}$ ) and O-$p_{xy}$ to the active band at the Fermi level. We perform  calculations 
for all relevant compounds within QS\emph{GW}+DMFT (marked as QS\emph{GW} in the table against respective compounds). We show the LDA results only for LCO as a comparision against the QS\emph{GW} results. Z, $\Gamma$, $N_{eff}$, T$_c^\mathrm{max}$ and $\Delta$ are extracted by post-processing the results from fully converged QS\emph{GW}+DMFT.}

\end{table*}





\bibliographystyle{prsty}

\end{document}